\documentclass[
]{ceurart}

\usepackage{enumitem}
\usepackage{xcolor}

\usepackage{listings}
\begin{document}

\copyrightyear{2023}
\copyrightclause{Copyright for this paper by its authors.
  Use permitted under Creative Commons License Attribution 4.0
  International (CC BY 4.0).}

\conference{}

\title{Insights from an OTTR-centric Ontology Engineering Methodology}

\author[1]{Moritz Blum}[%
orcid=0000-0003-4924-3903,
email=mblum@techfak.uni-bielefeld.de,
]
\cormark[1]
\fnmark[1]
\address[1]{Bielefeld University,
    CITEC, Inspiration 1, 33619, Bielefeld, Germany}

\author[1,2]{Basil Ell}[%
orcid=0000-0002-8863-3157,
email=bell@techfak.uni-bielefeld.de,
]
\fnmark[1]
\address[2]{University of Oslo, Problemveien 11, 0313 Oslo, Norway}

\author[1]{Philipp Cimiano}[%
orcid=0000-0002-4771-441X,
email=cimiano@techfak.uni-bielefeld.de,
]
\cortext[1]{Corresponding author.}
\fntext[1]{These authors contributed equally.}

\begin{abstract}
OTTR is a language for representing ontology modeling patterns, which enables to build ontologies or knowledge bases by instantiating templates. Thereby, 
particularities of the ontological representation language are hidden from the domain experts, and it enables ontology engineers to, to some extent, separate the processes of deciding about what information to model from deciding about how to model the information, e.\thinspace g., which design patterns to use. Certain decisions can thus be postponed for the benefit of focusing on one of these processes. To date, only few works on ontology engineering where ontology templates are applied are described in the literature.

In this paper, we outline our methodology and report findings from our ontology engineering activities in the domain of Material Science. In these activities, OTTR templates play a key role. Our ontology engineering process is bottom-up, as we begin modeling activities from existing data that is then, via templates, fed into a knowledge graph, and it is top-down, as we first focus on which data to model and postpone the decision of how to model the data.

We find, among other things, that OTTR templates are especially useful as a means of communication with domain experts. Furthermore, we find that because OTTR templates encapsulate modeling decisions, the engineering process becomes flexible, meaning that design decisions can be changed at little cost.

%{\color{green}
%In this paper we describe challenges that we have encountered, such as that the ontology cannot always be deduced from a template library, but only from a template library in combination with template instantiations.
%} % end green

\end{abstract}


\maketitle

\section{Introduction}

%\textbf{Model Driven Engineering}
%
%Model driven engineering is about defining a model that help to define and to give answers of the system under study without the need to consider it directly. (Model-driven engineering: A survey supported by the unified conceptual model - da Silva)
%=> template signatures also help to specify the ontology (model of a model)
%
%Model Driven Engineering has two main components:
%* Domain-specific modelling languages to describe a model
%* Transformation engines has to analyse the model and then synthesise the modelled system 
%(Model-Driven Engineering - Schmidt)
%
%OTTR templates are our domain-specific language to describe the model. 
%
%The templates signatures are transformed into the ontology. Instead to classical MDE, this does not have to be done is one step, instead, multiple nested/sub-templates can be created, until the structure is defined clearly.  We to model-to-model transformations until we to model-to-text (in our case model-to-ontology) transformations. (Model-driven engineering: A survey supported by the unified conceptual model - da Silva)

Knowledge representation through ontologies plays a crucial role across multiple domains. However, creating large ontologies poses notable challenges. %, and existing methodologies have to improve in addressing these complexities. 
In particular, multiple design decisions must be made during the ontology development process, e.\thinspace g., determining the relevant information to model and deciding on suitable ontologies or design patterns to build on. Additionally, collaborating with domain experts unfamiliar with the technical details of RDF or OWL further compounds the challenges faced in ontology development.

Just as modularization enabled the development of large software systems, this technique has found application in ontology engineering as well (see, e.\thinspace g., \cite{hannou2022acimov}). Ontology templates embrace the concept of modularization, facilitating the construction of ontologies through the instantiation of templates.
The type of modularization we realize is not that the content of an ontology is distributed into specific modules. Instead, in our case, templates encapsulate modeling decisions so that these modeling decisions do not need to be specified repeatedly and redundantly, which could lead to inconsistencies when the ontology evolves, but lead to uniformly modeled data.
Despite the potential advantages of ontology templates and template instances, so far, no generally applicable ontology engineering methodology
has made use of ontology templates as central design building blocks. 

%\textbf{TODO our contribution is ....}

We propose an ontology engineering methodology that is centered around Reasonable Ontology Templates~(OTTR)~\cite{skjaeveland2019pattern}. 
%We propose to improve ontology engineering by using Reasonable Ontology Templates~(OTTR)~\cite{skjaeveland2019pattern}. 
OTTR is a language and framework that introduces ontology templates, allowing to develop ontologies at a higher level of abstraction. The templates can be instantiated to construct ontologies. 

OTTR allows to decouple the process of determining the information to be modeled from the process of deciding how to model it, thus allowing a more agile development approach. We propose to leverage this feature to postpone design decisions until the data is pre-structured. Additionally, the abstraction through templates simplifies the communication with domain exerts by staying on a higher level without domain experts getting in touch with RDF.

We propose bottom-up ontology engineering and begin with existing data to start building the OTTR template headers. Domain experts actively participate and contribute their knowledge, which serves as a foundation for a discussion. 

Subsequently, we adopt a top-down refinement approach, utilizing OTTR template headers as a starting point to develop OTTR template bodies iteratively. Thereby, the ontology content is continuously refined by establishing sub-templates that model certain parts of the data. %{\color{green} Notably, domain experts do not have to search for suitable ontologies for reuse, instead, this will be done iteratively and naturally by ontology engineers.} % end green

The template-based ontology engineering approach is related to the model-driven engineering methodology (MDE), in that a model is generated from a meta-model. In our case, the meta-model is a set of template definitions and instances and the generated model is an ontology.

In the remainder of this paper, we present our methodology by outlining the steps to take, which do not have to be followed strictly linearly. Additionally, we discuss different design decisions, like when to split a template into multiple templates. %{\color{green}, challenges, such as deducing the ontology from the template library, and discuss remaining questions}. 
Moreover, we report our experience gained while developing a material science domain ontology, and we outline how we interact with domain experts and how we integrate real-world data to populate the ontology.

\section{Related Work}
In this paper, we focus on ontology engineering activities that make use of ontology templates.
As we are not proposing a genuinely new ontology engineering methodology, we do not provide a comprehensive overview of existing ontology engineering methodologies.

Existing ontology engineering methodologies include the Cyc methodology~\cite{cyc},
ONIONS~\cite{gangemi1996onions}, Methontology~\cite{fernandez1997methontology}, CommonKADS~\cite{schreiber2000knowledge}, the On-To-Knowledge Methodology (OTKM)~\cite{sure2004knowledge}, UPON~\cite{de2005proposal}, DILIGENT~\cite{Pinto2006}, DOGMA~\cite{jarrar2009ontology}, and the NeOn methodology~\cite{suarez2011neon}. Some of these methodologies are described and compared in a survey by
Iqbal et al.~\cite{iqbal2013analysis} and in a book by Maria Keet~\cite{keet2018introduction}. A survey by Simperl et al. focuses on collaborative ontology engineering methodologies~\cite{simperl2014collaborative} such as DILIGENT. Complementing these methodologies that introduce process models, there exist a wide range of ontology evaluation methodologies (see Vrande{\v{c}}i{\'c}~\cite{vrandevcic2009ontology} for an overview), such as the OntoClean~\cite{ontoclean} and a list of OWL ontology antipatterns~\cite{roussey2009catalogue}.

All of these methodologies have been developed before ontology templates were introduced. However, ontology templates can be seen as enabling ontology experts to develop domain specific languages (DSL). Thus, related work exists regarding design guidelines for DSLs~\cite{karsai2014design}.
We refer to the book by Martin Fowler~\cite{fowler2010domain} for a general introduction to DSLs. Mernik et al.~\cite{mernik2005and} discuss when and how DSLs should be used. In the context of ontology engineering, an expressive and redundancy-free DSL could be developed that is understandable by domain experts and abstracts from the particularities of the ontological representation language (i.\thinspace e., RDF(S) and OWL). 
E.\thinspace g., Semantic Application Design Language~(SADL)~\citet{crapo2013toward} offers a controlled-English language that translates into OWL or SPARQL.
In contrast to a DSL,
not every OTTR template needs to be domain specific. Instead, ontological modeling is modularized such that there are user-facing domain-specific templates which are understandable by domain experts, but these templates make use of modeling-encapsulating templates that step-by-step guide the translation of data into an ontology.

The idea behind Ontology Design Patterns (ODP) is to have available components \textit{that will support ontology design at the level that is more natural to domain experts and laymen, i.e. the level at which small, expertise-aware components can be assembled as easy-to-apply,
easy-to-customize building blocks}~\cite{gangemi2009ontology}.
According to Skj{\ae}veland et al.~\cite{skjaeveland2019pattern}, OTTR templates have a similar ambition, but go one step further by enhancing the practicality of applying ODPs and ensuring their consistent implementation. %Furthermore, the details of the ontology language (e.\thinspace g., OWL) and the data model (i.e., RDF) are hidden from the end user, thus providing a better level of abstraction.
Furthermore, OTTR templates
abstract from the particularities of the ontological representation language. 

%{\color{green}
%However, ODP and OTTR can be seen as compatible. OTTR templates can make use of ODPs, so that instantiating an OTTR template can mean reusing an ODP.
%}

Creating an ontology via OTTR template instantiation can be seen as a model-driven design method. Some works exist on model-driven ontology engineering~\cite{gavsevic2006model,pan2006model,cranefield2007bridging} and make use of the Ontology Definition Metamodel (ODM), developed by the Object Management Group (OMG). ODM enables to translate OWL ontologies into UML and back, but it does not support encapsulation of modeling decisions, as it is the case with OTTR templates.

\citet{lupp2020template} describe an ontology engineering methodology in which OTTR template headers are created from highly structured data by creating parameters that correspond to columns in tables. Therefore, this methodology focuses on a scenario where OTTR templates can be derived directly from the data. Moreover, they encounter a low number of data sources and no relations between those.

\section{Preliminaries: OTTR}
In this section, we provide a very brief introduction to Reasonable Ontology Templates (OTTR)~\cite{skjaeveland2019pattern}, which is \textit{a language with supporting tools for representing and instantiating RDF graph and OWL ontology modeling patterns.}\footnote{See \url{http://ottr.xyz}}

\begin{figure}[ht]
\begin{center}
% https://docs.google.com/presentation/d/1hdXespch_XCPEZVbymJfhHRBF9L8minkzIZDcpWaWy4/edit?usp=sharing
a \includegraphics[height=4cm]{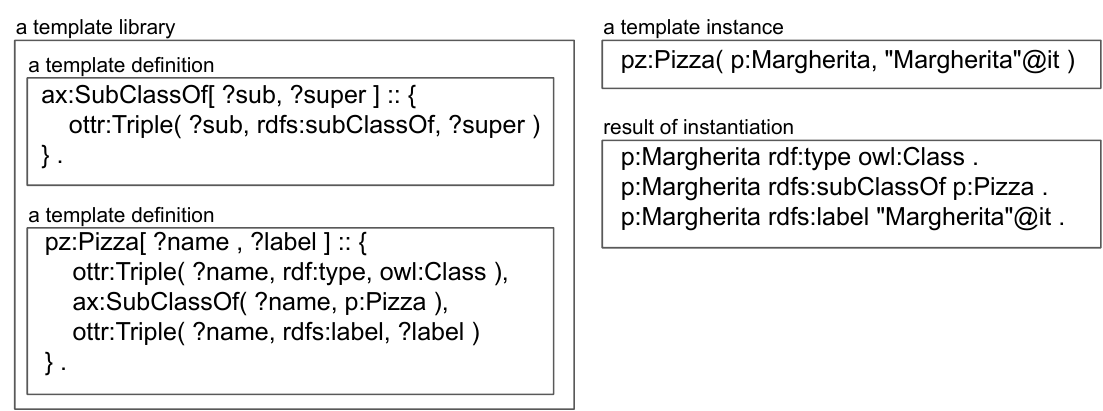}
\end{center}
\caption{
A pizza template library is shown on the left and an instance of the pizza template is shown on the right. The terms \texttt{p:Margherita} and \texttt{"Margherita"@en} are bound to the template parameters \texttt{?name} and \texttt{?label} to create an instance of the pizza template, whose triples are shown underneath. 
}
\label{fig_ottr_template_lib}
\end{figure}

An OTTR template library is a set of OTTR template definitions. See Fig.~\ref{fig_ottr_template_lib} for an example. Each template definition consists of a header (e.\thinspace g., \texttt{ax:SubClassOf[?sub, ?super]} -- \texttt{ax:SubClassOf} is the template's name; and  \texttt{?sub} and \texttt{?super} are its parameters) and a body (enclosed by curly brackets). Template bodies can refer to other templates -- thus, templates can be nested. \texttt{ottr:Triple} is the base template that creates RDF triples. A template instantiation denotes binding concrete values to the template parameters and to create triples. 

In this paper we distinguish between \textit{template instantiation}, i.e., the process of instantiating a template, and \textit{template instance}, i.e., the result of instantiating a template. Furthermore, we distinguish between templates that are only instantiated from within other templates (such as \texttt{ax:SubClassOf} in Fig.~\ref{fig_ottr_template_lib}), and those that are not (such as \texttt{pz:Pizza} in Fig.~\ref{fig_ottr_template_lib}). The latter we call user-facing templates. User-facing templates are instantiated by domain experts, or these template instances are generated by some tool.

We omit explaining parameter types and how type consistency is checked, optional parameters and default parameter values, lists, list expansion methods, and the different serialisation formats.

\section{OTTR-centric Ontology Engineering Methodology}
\label{methodology}

This section presents a list of steps and guidelines to develop an ontology using OTTR templates. It is important to note that this list of steps does not follow a strictly linear process model. 
%{\color{green}; instead, specific steps may be performed iteratively, and occasional backward steps might be required}. 
In addition, we discuss our practical experience after most steps, %. This part of our paper focuses on the practical insights, 
gained in the context of the DiProMag project.\footnote{See \url{https://www.dipromag.de}} % about applying the proposed OTTR-centric ontology engineering methodology. 
One of the ambitions of DiProMag is to model experiments related to materials, specifically magnetocaloric alloys, from the production, over the characterization to the prototypical application in an application ontology. 

From the ontology engineering perspective, the project context favors a bottom-up (from data to ontology) ontology engineering approach, as properties of materials are measured by machines automatically, producing digital data that largely adheres to common data formats and standards. As a result, the structured data can be utilized in designing the template headers. 

The subsequent steps to be followed are as follows: define the scope of the ontology and screen the available data~(Sec.~\ref{scope}), design of template headers~(Sec.~\ref{header}), template header design verification and documentation~(Sec.~\ref{header_verification}), design of template bodies~(Sec.~\ref{body}), handling of axiomatic triples~(Sec.~\ref{axioms}),  template body documentation~(Sec.~\ref{body_doc}), template library documentation~(Sec.~\ref{lib_doc}), and template instantiation \& data integration~(Sec.~\ref{instantiation}).

The ontology engineering process involves close collaboration between ontology engineers and domain experts. The ontology engineers primarily manage most activities based on the insights and feedback provided by domain experts during various discussions. However, there are specific tasks that ontology engineers handle independently, such as identifying suitable existing ontologies and ODPs. Conversely, domain experts take the lead in documenting the process for inputting data into the templates, as it heavily relies on domain-specific knowledge. 

Although developing OTTR templates requires ontology experts to possess additional skills, in the long run making use of OTTR templates simplifies ontology engineering, population, and maintenance.

%{\color{green} The OTTR templates created do not add an overhead, as they formalize design decisions that have to be made in any methodology. Instead, they improve, e.\thinspace g., the communication between ontology engineers and domain experts. Moreover, the OTTR templates can be used to populate and extend ontologies easily through instantiations.}

%most activities: managed by ontology engineers, requiring input from domain experts.  \\
%some activities: are taken care of by ontology engineers only, e.g. reuse of existing ontologies and ODPs.\\
%some activities: are taken care of by domain experts only, e.g. documentation about how to put in data into the templates as this is very domain dependent

%note that the list below is not a sequence of steps. a step can be carried out repeatedly. we here do not define a (linear) process model.

\subsection{Define the Scope of the Ontology and Screen the Available Data}
\label{scope}

In the first step, we follow the common practice of ontology engineering: we define the ontology's scope through competency questions~(CQs)~\cite{gruninger1995methodology, presutti2009extreme} and let domain experts examine relevant data (e.\thinspace g., datasets, records, and other information repositories) with those CQs. Based on this foundation, the ontology engineers engage in a discussion with the domain experts so that the ontology engineers understand the domain and the data's meaning, as well as its structure and parameters.

%The first step in ontology engineering is to define the ontology's scope and specify the direction for the entire development process. Similar to other ontology engineering methodologies, we start by specifying competency questions – key inquiries that users seek answers to by querying the ontology to ensure that the ontology will match its users' needs.

%Guided by these competency questions, domain experts will screen the available data, which involves an examination of available datasets, records, and other information repositories. 

%The collected data is then presented to and discussed with the ontology engineers so that the ontology engineers understand the domain and the data's meaning, as well as its structure and parameters. These discussions should be documented informally by taking notes, as this information will be crucial in most of the following steps. 

As in other methodologies, there is a risk of overlooking relevant information or the risk of modeling information irrelevant for the CQs. For example, modeling all sensor readings over time might be unnecessary if the CQs only ask for certain values. Therefore, this should be discussed with the domain experts with respect to the CQs which help to find missing source data and help to identify and filter out irrelevant data. 
%Also, the detail level must be considered to optimize efficiency and avoid undue complexities. For example, modeling all sensor reading in the ontology might be unnecessary if the competency questions only ask for specific aggregated values. Therefore, the ontology should only capture these aggregated values.
%Second, as collecting all related data can lead to a large amount of data, one should keep track of the alignment with the ontology's purpose. The competency questions can serve again as a guide, helping to identify and filter out irrelevant data in deep discussion between domain experts and ontology engineers asking critical questions. Also, the detail level must be considered to optimize efficiency and avoid undue complexities. For example, modeling all sensor reading in the ontology might be unnecessary if the competency questions only ask for specific aggregated values. Therefore, the ontology should only capture these aggregated values.
Throughout the process, the CQs may evolve, being adapted and refined as domain experts continuously offer insights. %This iterative approach fosters continuous improvement, culminating in a robust and purpose-driven ontology that effectively represents the domain and meets the users' needs.

\paragraph{Experiences:} 
Our CQs describe knowledge that would also be published in research papers, like results of experiments, but also details required for their reproduction. In addition, the ontology should be usable by machine-learning components to generate hypotheses about the properties of materials. For this reason, and to make the data easily queryable with SPARQL, it is essential to follow standards as much as possible, e.\thinspace g., using the International System of Units, and to model the data uniformly. In general, our CQs are not designed specifically for our OTTR template based ontology engineering methodology, and, therefore, do not differ from CQs developed in other methodologies.

During data collection, we noticed a substantial degree of heterogeneity in the data management practices applied to the data. To facilitate future data integration, we not only collected representative data samples but also metadata like where the data is located (i.e., machine, volume, folder, etc.), how it can be accessed (i.\thinspace e., file, database, service), how often it is updated, and the format of the files.

\subsection{Design of Template Headers}
\label{header}

%In this step, the ontology engineers decide which data is handled by which templates. 
%When designing headers for the data that already exists and the data that is expected to exist in the future, 

% Furthermore, we try to avoid modeling redundancy. That means that we do not create template headers and templates in general in which similar kinds of data are handled in a similar way. Even though initially the templates might model the data in the same way, an ontology engineer could change the modeling of one template without changing the modeling of the other template, which would result in non-uniform modeling of similar kinds of data.

%often JSON, complex tables
%data changes nicht einfach

%same data put in multiple instantiations. not ideal.
%modeling reducnancz

%do we need to distinguish between (CSV) table vs. relational database vs. whatever? ideally not. maybe we focus on tables.
%the most simple approach would be: for each table, create a header that contains a parameter for each column.

%the table design might not be ideal. this would lead to the following issues. if X is the case, then one designs headers in a way that only a subset of columns are taken into account. this decision is based on Y.

%this might be related to database normalization and functional dependencies. we don't need to go into too much detail, but at the moment I need to understand more.

%the main things to consider when creating the template headers: 

% intro: it's not that simple
After collecting data relevant to the design of the ontology, OTTR template headers can be developed. An OTTR template header mainly specifies a template by giving it a name and listing its parameters together with their data types. The abstraction through template headers simplifies the communication with domain exerts by discussing on the level of template names and parameters while abstracting from how the data is modeled in RDF. 
When designing template headers, trade-offs need to be made, taking the following principles into account:
\begin{itemize}
 \item \textbf{Complexity of templates vs. complexity of template relations} 
 If a template (header) has many parameters, then it can become difficult to use or update the template (header). Thus, it can be advisable to split a template header into multiple template headers while at the same time not distributing parameters that form a group over multiple templates. However, distributing information across multiple templates can also increase complexity for end users, as they need to understand how these templates are related so that they can be used correctly. The so-called workflows, that we describe in Section~\ref{header_verification}, are intended to help here.

% In some cases where the complexity of templates gets too large, a factorization into sub-templates is advisable. 
%    However, splitting information across multiple templates can also increase complexity for end users, as there are more relations between templates that might have to be established such that a connected ontology is build. Therefore, we  had to provide so-called workflow we introduce in Section~\ref{header_verification}.

\item \textbf{Conceptual similarity} Data that is conceptually similar should be taken care of by the same template. In principle, one can follow standard practices from relational modeling. A template instance can be seen as a tuple in a relational model. Thus, design principles for entities and relationships from entity-relationship-modeling can be used when designing template headers~\cite{10.1145/320434.320440}. %One difference, though, is the ambition to reduce redundancies in relational models by making use of functional dependencies, as redundancies can lead to inconsistencies. In the context of 

\item \textbf{Multiple similar but specific templates vs. few general templates} For a set of similar situations, either for each situation another template is created such that each of these templates almost models the same data. Thus, each template is specific to one situation. Alternatively, one template is created that fits multiple situations, where in each situation not all situation-specific parameters need to be used. For domain experts, situation-specific templates might be easier to understand and use. For ontology engineers, sets of situation-specific templates require more care to ensure uniform modeling across situations.

\item \textbf{Avoid duplication wisely, but not at all costs}
This principle from software engineering applies to template design, too. Although similar situations should be modeled similarly, which could mean  with the same template, it can be that the similarity might be likely to disappear in the future when the ontology evolves. Here, the single-responsibility principle needs to be taken into account.
%In the case, having separate templates for similar things can be beneficial.

%principle from SE. keep things separate, although they look similar, are subject to different processes

\item \textbf{Avoiding data redundancies and data inconsistencies} 
Data redundancy can occur at the instantiation level and at the output level. The former means, that the same information is used in multiple template instances, e.\thinspace g., 
in instances of different templates or of the same template. This can lead to inconsistencies, if the template instances need to reflect externally available data that can change, as changes in the external data require instances to be updated and in that case, one needs to identify all instances that need to be updated. %Otherwise, this leads to data inconsistencies. %Note that having multiple templates with similar parameters does not automatically result in data redundancy at the instantiation level.

Data redundancy at the output level means that a triple or a set of triples is generated by more than one template instance. As an RDF graph is a set of triples, these redundant triples do not require additional memory. However, if an ontology engineer wishes to make changes such that a specific triple does not anymore occur in the output, then identifying all template instances that need to be adapted might be challenging.

\item \textbf{Additional Metadata} Beyond modeling parameters for existing data, e.\thinspace g., from files or databases, the ontology engineer can add further parameters to the template headers so that provenance information or any other meta data about the data generating process can be captured. 

%about data-generating process
%Another important topic that should be regarded in the template header development is metadata, such as provenance information. This ensures that templates accurately capture data origin and source details.

%\item \textbf{Redundancy} Multiple templates model the same data. This is in general not a modelling issue as sub-templates can ensure a uniform modelling and redundant triples would be removed in the RDF graph anyway. It is important to ensure throughout the development process that not multiple templates model similar information differently. However, in terms of data integration and maintenance, duplicated modelling can cause issues if some triples should be removed from the resulting RDF graph, which are modelled by multiple templates.  
    
\end{itemize}

To maintain clarity and consistency throughout the template design, we document dependencies between parameters within one template and between templates. This practice will ensure appropriate modeling of the template bodies.

The result is an initial collection of template headers. It might be useful to improve them iteratively in collaboration with domain experts.% This optimization process can help to achieve three primary objectives: data coverage, minimal redundancy in the stored information across templates. This might also require refactoring or establishing connections between templates.

\paragraph{Experiences:} 
% selection of first template candidates
As potential template candidates, we used one template per process step, i.\thinspace e., one template per material type, one template per synthesis method, and one template per measurement method. The logical relation between those steps served as a basis for connecting the templates.

%what about JSON etc,? not as simple as having a table, create a header with one parameter per column.

% I think the goals and ambitions do not really match our story...
%and a template for goals and ambitions.

% selection of first template parameters in order to create first template headers
Next, we derived the template parameters directly from the available data. In cases where there is no data available to answer a certain CQ, the domain experts provided us with some artificially created data similar to the expected real data. This process directly involved domain experts that supported us with details about the different parameters and their data types. 

% iterative process of adjusting the template headers 

The resulting template headers were iteratively improved to follow the introduced design principles. 
For example, one critical aspect of splitting data across multiple templates concerns the correlations and dependencies between data points. Although such connections may not be explicitly present in the raw data, modeling them explicitly in the ontology might be essential. E.\thinspace g., two machines are used to measure a material's physical properties and the experiment's environmental conditions. Simply considering the data files, both appear to be separate and unrelated. However, the physical properties might only be meaningful in the context of the particular environment in which they were measured.

\subsection{Template Header Design Verification and Documentation}
\label{header_verification}

%hier geht es nur um die, die user facing sind.

%story:
%template headers should be verified before template bodies are created.

%verification and documentation differs depending on whether the template is or is not end-user facing.

%end-user facing templates are verified by letting domain experts instantiate them with realistic or real data
%instantiating can mean to fill out a form
%this has the benefits that ...
%by letting domain experts fill out form it could be found that ... (or, maybe these are already our experiences.)

%the domain experts would not necessarily use datatypes. thus, ontology experts then need to make sure parameter constraints are met.

%once template headers are verified they need to be documented.
%documentation is done by domain experts, in supervision by ontology engineers. documentation means filling out a table? free text? a comment per field? note about the instantiation order?

%how are templates verified that are not end-user facing?
%probably in a similar way? however, one could have templates with empty body and have domain experts input realistic data. then, also types and other constraints can be checked.

%note: a filled out template, even if it is just a form in an office document, is a relevant asset to document a template.

%---

Verifying the developed template headers is essential at this stage, as issues will propagate further to the development of the template bodies. To verify the headers of user-facing templates, domain experts can instantiate templates with sample data by informally filling out forms or tables where they do not need to focus on syntax and specific data types. As a result, design decisions from the header design~(Sec.~\ref{header}) might be revisited, because, e.\thinspace g., some data structures (like complex JSON objects or tuples) are not represented correctly by the template parameters. 

% During template instantiation, one could also observe whether parameters may need to be split into multiple parameters or must be processed beforehand, e.\thinspace g., aggregation of multiple values or conversion like a translation. For example, tuples must be split into multiple parameters as accessing the elements at certain positions is not possible in the template body, or an aggregation of sensor values could compute the mean values of those, because the individual values will never be used. 

To simplify the instantiation of templates with (multiple) relationships to other templates, we propose to define workflows that describe how templates are related to each other. These workflows provide instructions for the instantiation order to build one single connected ontology instead of multiple unconnected ontologies which model certain parts of the information separately. Having such a proposed instantiation order has two benefits. First, domain experts have a clearly-defined sequence of steps to follow, which reduces the chance of forgetting a step. Second, with an instantiation order that ensures that the ontology is connected after each instantiation, inspecting and querying the ontology while working through a sequence of instantiations can become easier.

After creating the template headers, domain experts document user-facing templates to ensure easy understanding and usability by other domain experts. Ontology engineers only provide support on topics like the data types or the naming conventions. 

The resulting documentation includes a general description about the purpose of the template~(what data can be modelled; answering questions about when and how the template is applied), modeling limitations (explaining what is not modeled or what can not be modelled), and specific details like data types, the data a parameter name refers to, default values, and examples of parameter values. Moreover, the documentation provides valuable information for the development of the template bodies, e.\thinspace g., the information about default values and example parameter values can serve as information to build a first taxonomy, and will ensure that domain knowledge is modelled correctly.

\paragraph{Experiences:}
% validation of OTTR headers & user interface

We developed \textit{OTTR Extension},\footnote{See \url{https://www.mediawiki.org/wiki/Extension:OttrParser}} an extension for Semantic MediaWiki~(SMW),\footnote{See \url{https://www.semantic-mediawiki.org/wiki/Semantic_MediaWiki}} which enables to use OTTR functionality within SMW. The \textit{OTTR Extension} enables to define and instantiate OTTR templates through SMW's web interface and can generate convenient input forms for template headers automatically. We used this SMW to validate the user-facing template headers through a manual instantiation by the domain experts. As a result, we recognized minor issues with the developed template headers, e.\thinspace g., a parameter was defined as a single value, whereas a list was necessary.

% documentation
For the documentation of template headers, we encouraged documenting the parameters in tables~(i.e., parameter name, data type, default values, example values, and description). % {\color{green} for a structured overview}.

Furthermore, workflows recommending an instantiation order were developed based on the template dependencies to avoid mistakes when connecting templates through URIs. E.\thinspace g., a material should be specified conceptually before modeling its measured properties.

\subsection{Design of Template Bodies}
\label{body}

Based on the template headers that define the shape of the input data and based on the provided documentation, the corresponding template bodies are developed by the ontology engineers. 

As by other ontology engineering methodologies propose, we encourage the reuse of existing ontologies with approved modeling decisions inside the template bodies to ensure interoperability with other ontologies~\cite{noy2001ontology}. Ontology engineers make use of the template header documentation when identifying relevant ontologies and ODPs.

The development of the template bodies is straightforward, as a pre-structuring is done through developing template definitions and through the ontologies that will be reused. Here, we refer to established ontology engineering methodologies that can be used to develop the ontology inside the template bodies. 

The major decisions of the ontology engineer with respect to OTTR are to create appropriate sub-templates. As our approach works top-down, templates with many parameters are the starting point to iteratively develop sub-templates.  %{\color{green}Creating a clear separation between the two sets of templates is important to ensure the template library is easy to use and maintain.} 

As we consider our proposed methodology to be non-linear, steps back are always possible. Nevertheless, the template headers usually do not require significant adjustment caused by issues encountered during the development of the bodies. The only conflict might be the incompatibility of the developed template library to the ontologies selected for reuse. An example is that a reused mid-level ontology defines a process that can not be modelled in a single template. Instead, the various steps are spread across multiple templates. In those cases, it must be discussed with the domain experts whether it makes sense to adjust the template headers or to use different ontologies and ODPs. %{\color{green}In the first case, the documentation has to get an update. }

Where for the design of template headers the avoidance of redundancy, specifically data redundancy~(see Section~\ref{header}), is an important design principle, for the design of template bodies a relevant design principle is the avoidance of modeling redundancy. With modeling redundancy, we mean that multiple template bodies implement similar design decisions. This is problematic, as this can lead to modeling inconsistencies. Therefore, modeling redundancy should be avoided wisely, but not at all costs, too, and appropriate sub-templates should be used instead.

It is important to note that domain experts do not need to understand the template bodies. Instead, we propose to show them graph visualizations of the template relations and seek their feedback on their sufficiency. The main requirement is that the ontology can be queried to answer the competency questions, which can be done in ontology validation similar to~\cite{BOOSHEHRI2021100074}.

\paragraph{Experiences:}
Drawing from mid-level ontologies and ODPs, we decided to build on the PMD core ontology\footnote{See \url{https://github.com/materialdigital/core-ontology/}} and EMMO.\footnote{See \url{https://github.com/emmo-repo/EMMO}} The development of the template bodies was straightforward and the use of sub-templates lead to a faster development process as similar data occurs frequently across our data which can be covered by the same sub-templates. 
%
%{\color{green}We will not go into detail about the concrete template body development here, as} 
A lot of the pre-structuring was done in previous steps, and the header documentation served as a good source of information. We observed that a collection of similar physical experiments often share the same data about the environment the experiments take place, like temperature and humidity. This hints at the fact that a sub-template should be introduced to model the (shared) data about the environment.

\subsection{Inclusion of Axiomatic Triples into the Ontology}
\label{axioms}

%When developing an ontology by designing and instantiating OTTR templates, the ontology, for example in the form of an RDF file, is an outcome of instantiating the templates. For example, for an axiom expressing that a given property is symmetric, there are several options to ensure that this axiom is contained in the resulting RDF file. We briefly discuss two disadvantageous methods before introducing the method we believe to be most suitable.

When developing an ontology by designing and instantiating OTTR templates, an RDF file is the result of instantiating the templates. If the resulting ontology should contain axioms, for example that a certain property is symmetric, then the question is where to define these axioms so that they are contained in the resulting output.

We propose the following method.
If a template body makes use of a term (i.e., the template body mentions it) about which we want to specify axioms, then the template's body instantiates another template that encapsulates all axiomatic triples about that term. Thus, that template is the only place where the axiomatic triples are stored, and an ontology engineer knows where to find these axioms. E.\thinspace g., the domain and range axioms for the property \textit{pz:hasTopping} would be added through a \textit{pz:AxiomHasTopping} template call. Moreover, these axioms can be defined through template calls, e.\thinspace g., \textit{o-owl-ma:DomainRange(pz:hasTopping, pz:Pizza, pz:Topping)}.\footnote{See \url{http://tpl.ottr.xyz/owl/macro/0.1/DomainRange}}

If a term is passed to a template for which no template exists that makes use of that term and could thus instantiate the template that generates the axioms about that term, then the above solution does not work. Instead, the solution is to develop templates that can then be instantiated about that term that then generate the axioms.

\subsection{Template Body Documentation}
\label{body_doc}
To ensure the reuse of templates and the maintenance of their bodies, we propose a template body documentation 
%{\color{green}including end-user-facing templates and sub-templates,} 
that is achieved through comments in the OTTR code. This documentation is technical and intended for the ontology engineers (not the domain experts) to explain the functionality and purpose of the statements inside the template bodies. 

Additionally, OTTR offers the DocTTR template library, designed to document OTTR templates using templates. The DocTTR\footnote{See \url{http://tpl.ottr.xyz/p/docttr/0.1/}} template library includes various documentation elements, including change notes, examples, parameters, provenance information, and versioning~\cite{skjaeveland2018practical}. 
%{\color{green}Therefore, well-established ontologies like Dublin Core (DC)\footnote{See \url{https://www.dublincore.org/resources/glossary/ontology/}}, PAV,\footnote{See \url{https://github.com/pav-ontology/pav}} and SKOS\footnote{See \url{https://www.w3.org/TR/2008/WD-skos-reference-20080829/skos.html}}, are used by DocTTR to facilitate consistent and standardized documentation practices.}

Overall, the template body documentation enhances collaboration and knowledge sharing.

%The template body documentation through comments in the OTTR code. Moreover, OTTR provides the Docttr template library for documenting ChangeNote, Deprecated, Example, Parameter, Provenance, header, Version that relies on, e.g. the Dublin Core (DC) ontology,  PAV ontology or the skos ontology. 

\subsection{Template Library Documentation}
\label{lib_doc}

The final documentation step has the goal to create an overview over the template library, including the structure, relationships, and functionality of the templates in the template library to the ontology engineers and domain experts. %{\color{green}The documentation can describe, e.\thinspace g., which templates are user-facing and which ones are not.}

The documentation process encompasses the creation of the following parts:
\begin{enumerate}

    \item \textbf{List of all Templates:}
 The documentation should list all templates in the library, showing which templates are user-facing and which ones are sub-templates and thus play an internal and supporting role within the library. 

    \item \textbf{Template Inclusion/Call Hierarchy:}
Understanding the template call hierarchy is crucial for understanding the dependencies among templates. This hierarchy can be derived automatically from the template library, but it is beneficial for both, ontology engineers and domain experts, to visualize and comment on it.

    \item \textbf{Instantiation Order and Naming Guidelines:}
Documenting the instantiation order is important to avoid mistakes when connecting templates. These instantiation orders must be recorded throughout the development process to provide a logical sequence for domain experts to follow. Additionally, guidelines for naming URIs are important to ensure consistency throughout the ontology. 

    \item \textbf{Individual Template Documentation:}
This denotes the documentation already provided in Section~\ref{header_verification} and Section~\ref{body_doc}. The documentation ensures that new users (domain experts \& ontology engineers) understand, and existing users can look up, each template's functionality, ensuring a correct utilization of the template library. 
\end{enumerate}

%{\color{green}To facilitate easy access and reference, the template library documentation should be a separate document, possibly in the form of dedicated Wiki pages or text documents to ensure that the documentation is accessible for both, ontology engineers and domain experts.}

By documenting the template library in the outlined way, ontology engineers and domain experts can efficiently find out which templates to instantiate for a given need and ontology engineers can efficiently maintain the template library in a structured and organized manner throughout its lifecycle. 

For the publication of the template library, we propose to use either a Wiki or GitHub. 

\paragraph{Experiences:}
In our project, we use a SMW to organize and store the template library documentation along with the templates. 
%{\color{green}Each template has its own page that holds the documentation associated with the template. Furthermore,} 
We create pages to address the aspects mentioned before, and we make use of the semantic functionalities of the SMW to make the template library and the documentation queryable, e.\thinspace g., to generate a list of all templates tagged to belong to a certain topic, project, or user. 

We introduced the naming conventions into the documentation, as we observed that the naming of URIs and templates held significant importance for the domain experts. It allowed them to easily identify information within the ontology and ensured reusability in other projects.

%This comprehensive template library documentation includes the previously described workflows of template instantiations, ensuring a clear understanding of the instantiation order. Templates that need to be instantiated first before others can be instantiated will be identified, along with guidelines for naming URIs to ensure consistency throughout the ontology.

%The documentation will also encompass all template details, including their purpose, parameters, and dependencies, ensuring that users understand each template's functionality entirely. 

%To facilitate easy access and reference of the template library, the documentation should be a separate document, possibly a Wiki or raw text documents. This approach ensures that the pattern library is well-documented, accessible, and a valuable resource for ontology engineers and domain experts.

%And the last documentation step is to document the pattern library as a whole.  It is important to build an overview about which templates are at least sometimes user-facing, which templates are never user-facing, what is the template inclusion/call hierarchy.

%This template library documentation will also include the previously described workflows of template instantiations and well as all template documentaitons
%Instantiation order: first instantiate this template before that template can be instantiate, name those URIs, ...

%Therefore, the documentation will also cover naming conventions for URIs. 

%documentation is a separate document, e.g. a Wiki or some simple text documents.

\subsection{Template Instantiation \& Data Integration}
\label{instantiation}
Based on the template library, the ontology is generated by instantiating the templates with data to populate the A-Box, the T-Box, or both. Templates allow to build the T-Box step by step in contrast to traditional ontology engineering methods, which encourage collecting many details of the T-Box, e.\thinspace g., to build the subclass hierarchy, before starting to populate the A-Box.

We have not yet explained how to integrate real-world data to instantiate templates and how to maintain the ontology. Therefore, a pipeline can automate the process of extracting, formatting, and feeding the data into OTTR templates (the information about the data sources collected, as described in Sec.~\ref{scope}, can be used) -- thus, generating template instances. Then, the Lutra engine\footnote{See \url{https://gitlab.com/ottr/lutra/lutra}} takes care of expanding the templates and to generate an ontology which can be stored in a triple store. 
This step is less about ontology engineering but more about data integration. Nevertheless, the data integration activities lead to templates instantiations being generated and templates being instantiated.
%{\color{green}It depends on the specific project whether the availability of new data triggers the generation of template instances, or whether some process regularly checks for new data and then triggers the generation of instantiations. Thus, it depends on the project's data ingestion architecture or an \textbf{ETL} process already in place.} 
New data can then reside in databases, as files, in the form of sensor readings etc. This process might involve aggregations of data, stream processing, complex event processing, synchronization, archiving of original data, dealing with data heterogeneity and quality issues, etc. Furthermore, it needs to be decided where the template instances should reside and where to store the results of the instantiations (i.e., in a triple store or as a file).

%one could mention that the wiki can contain example instantiations that create actual triples that can be queried and visualized. 

%{\color{green} Changes can happen on the input data or the template bodies. In order to update, insert, or remove triples in case of changes from the ontology, information about the underlying data of a template instantiation and the associated triples with an instantiation should be stores. Otherwise, a complete re-instantiation would be necessary.  }

\paragraph{Experiences:}
In our case, the templates are instantiated either from data, or manually by domain experts, e.\thinspace g., through an interface in SMW or in eLabFTW.\footnote{See \url{https://www.elabftw.net}}
 
In our case, eLabFTW generates template instances and sends these to our SMW (i.e., creates new wiki pages that then contain template instances), which returns unique identifiers for each template instance to allow changing the data later. If template bodies are changed by an ontology engineer, then the template bodies can be easily updated in SMW, and the triples resulting from instantiations will automatically update. Therefore, SMW serves as a platform for managing templates, their instances, and as a triple store.

Some template instances contain data obtained via experiments by the material science researchers involved in our project. It is important to distinguish between data that has already been reported in scientific publications and data that has not. In the latter case, this data should be kept internal. 
We used parameters to indicate the publication status. Our template library does not include sensitive data such as the template instances and can, therefore, be published.

\section{Discussion}
We would like to use the discussion to address two open questions regarding the use of ODPs in combination with OTTR and the extraction of the ontology from the templates definitions. 

%Martin~\cite{skjaeveland2019pattern} already describes that ODPs can be used inside of OTTR templates. 

\paragraph{ODPs in combination with OTTR}
There are two ways of using OPDs in combination with our OTTR-centric ontology engineering methodology. The first method is to have a template that only instantiates a single ODP. This way, the ODP can be used in other templates easily. Then, the OTTR features like, e.\thinspace g., list expansion, dealing with optional/default parameters, and checking (datatype and type) constraints, make using ODPs much easier than applying them manually. However, depending on the complexity of the ODP and the template library, this method is not always feasible. Therefore, the second method is to distribute an ODP across multiple templates. In this case, it is necessary to introduce new relations between templates such that the ODP is still followed. 

\paragraph{Extraction of the ontology}
After the development of the OTTR template library, the question raises how to extract the ontology from the template definitions only. We recognize that the ontology is generated from a combination of the template library and the template instances. Therefore, extracting the complete T-Box from the template library alone is not a straightforward task, and it might be the case that the entire T-Box can only be obtained when also incorporating the template instances.

%is in the most cases first possible after instantiations have taken place, as the ontology might be incomplete before that point.

\section{Limitations \& Conclusion}
In this paper, we outline an ontology engineering methodology that relies on OTTR templates for representing ontology modeling patterns. We present the key steps that include defining the ontology's scope, designing template headers, verifying and documenting them, incorporating axiomatic triples, creating and documenting template bodies, as well as documenting the entire template library, and instantiating templates and integrating data.

Additionally, we share experiences from a project, demonstrating how OTTR templates naturally support ontology engineering. We observed benefits such as flexibility in the case where modeling decisions are re-engineered and enhanced communication with domain experts. 

What we propose in this paper is not a comprehensive ontology engineering methodology, %nor are we in a stage where we have collected broad empirical evidence about its advantages and limitations.
because it should be combined with existing ontology engineering practices, e.\thinspace g., for the development of competency questions, the creation of the template bodies, and for the evaluation of ontologies. %\textbf{ (see e.\thinspace g., Vrande{\v{c}}i{\'c} \cite{vrandevcic2009ontology})}. 
The direction we are exploring is towards incorporating principles of modularization into the design of ontologies so that design decisions are not represented redundantly. Much research has been carried out on abstraction and modularization in the domain of software engineering, and we believe that there are multiple principles and techniques, from model-driven engineering and beyond, that could be adapted and applied to the field of ontology engineering.

%\bigskip

%\textbf{TODO} remove page numbers and page headers before submission

%\textbf{TODO} shorten to 12 pages

%\textbf{TODO} fix remaining latex warnings, if possible

%\textbf{TODO} makse sure that "template instantiation" and "template instance" are used correctly:\textbf{ instantiation = template head + values; instance = ontology/triples obtain through an instantiation }

%\textbf{TODO} "experience" sounds like what worked well and what didn't. often, that is not what we describe, at least not explicitly. see e.g. the part about SMW. maybe, what we describe is how we realized that step and we what we described is what has worked well for us. \textbf{Something like implementation?}

%\textbf{TODO} Ende section 4.3

%\textbf{TODO} shorten red text

\begin{acknowledgments}
This work was done in the context of DiProMag, a BMBF (German Federal Ministry of Education and Research) funded research project under Grant No. 13XP5120B, and Sirius with the Norwegian Research Council project No. 237898.
\end{acknowledgments}

\bibliography{main}

\begin{thebibliography}{32}
\expandafter\ifx\csname natexlab\endcsname\relax\def\natexlab#1{#1}\fi
\providecommand{\url}[1]{\texttt{#1}}
\providecommand{\href}[2]{#2}
\providecommand{\path}[1]{#1}
\providecommand{\DOIprefix}{doi:}
\providecommand{\ArXivprefix}{arXiv:}
\providecommand{\URLprefix}{URL: }
\providecommand{\Pubmedprefix}{pmid:}
\providecommand{\doi}[1]{\href{http://dx.doi.org/#1}{\path{#1}}}
\providecommand{\Pubmed}[1]{\href{pmid:#1}{\path{#1}}}
\providecommand{\bibinfo}[2]{#2}
\ifx\xfnm\relax \def\xfnm[#1]{\unskip,\space#1}\fi
%Type = Inproceedings
\bibitem[{Hannou et~al.(2023)Hannou, Charpenay, Lefran{\c{c}}ois, Roussey,
  Zimmermann, and Gandon}]{hannou2022acimov}
\bibinfo{author}{F.-Z. Hannou}, \bibinfo{author}{V.~Charpenay},
  \bibinfo{author}{M.~Lefran{\c{c}}ois}, \bibinfo{author}{C.~Roussey},
  \bibinfo{author}{A.~Zimmermann}, \bibinfo{author}{F.~Gandon},
\newblock \bibinfo{title}{{The ACIMOV Methodology: Agile and Continuous
  Integration for Modular Ontologies and Vocabularies}},
\newblock in: \bibinfo{booktitle}{2nd Workshop on Modular Knowledge},
  \bibinfo{year}{2023}, pp. \bibinfo{pages}{1--13}.
%Type = Inproceedings
\bibitem[{Skj{\ae}veland et~al.(2018)Skj{\ae}veland, Lupp, Karlsen, and
  Forssell}]{skjaeveland2019pattern}
\bibinfo{author}{M.~G. Skj{\ae}veland}, \bibinfo{author}{D.~P. Lupp},
  \bibinfo{author}{L.~H. Karlsen}, \bibinfo{author}{H.~Forssell},
\newblock \bibinfo{title}{Practical ontology pattern instantiation, discovery,
  and maintenance with reasonable ontology templates},
\newblock in: \bibinfo{booktitle}{The Semantic Web--ISWC 2018: 17th
  International Semantic Web Conference, Monterey, CA, USA, October 8--12,
  2018, Proceedings, Part I 17}, \bibinfo{organization}{Springer},
  \bibinfo{year}{2018}, pp. \bibinfo{pages}{477--494}.
%Type = Article
\bibitem[{Elkan and Greiner(1993)}]{cyc}
\bibinfo{author}{C.~Elkan}, \bibinfo{author}{R.~Greiner},
\newblock \bibinfo{title}{{Building Large Knowledge-Based Systems:
  Representation and Inference in the Cyc Project: D.B. Lenat and R.V. Guha}},
\newblock \bibinfo{journal}{Artificial Intelligence} \bibinfo{volume}{61}
  (\bibinfo{year}{1993}) \bibinfo{pages}{41--52}.
%Type = Inproceedings
\bibitem[{Gangemi et~al.(1996)Gangemi, Steve, and
  Giacomelli}]{gangemi1996onions}
\bibinfo{author}{A.~Gangemi}, \bibinfo{author}{G.~Steve},
  \bibinfo{author}{F.~Giacomelli},
\newblock \bibinfo{title}{{ONIONS: An ontological methodology for taxonomic
  knowledge integration}},
\newblock in: \bibinfo{booktitle}{ECAI Workshop on Ontological Engineering},
  volume~\bibinfo{volume}{95}, \bibinfo{year}{1996}, pp.
  \bibinfo{pages}{95--106}.
%Type = Inproceedings
\bibitem[{Fern{\'a}ndez-L{\'o}pez et~al.(1997)Fern{\'a}ndez-L{\'o}pez,
  G{\'o}mez-P{\'e}rez, and Juristo}]{fernandez1997methontology}
\bibinfo{author}{M.~Fern{\'a}ndez-L{\'o}pez},
  \bibinfo{author}{A.~G{\'o}mez-P{\'e}rez}, \bibinfo{author}{N.~Juristo},
\newblock \bibinfo{title}{{Methontology: from ontological art towards
  ontological engineering}},
\newblock in: \bibinfo{booktitle}{1997 AAAI Spring Symposium},
  \bibinfo{year}{1997}, pp. \bibinfo{pages}{33--40}.
%Type = Book
\bibitem[{Schreiber(2000)}]{schreiber2000knowledge}
\bibinfo{author}{G.~Schreiber}, \bibinfo{title}{{Knowledge engineering and
  management: the CommonKADS methodology}}, \bibinfo{publisher}{MIT press},
  \bibinfo{year}{2000}.
%Type = Inbook
\bibitem[{Sure et~al.(2004)Sure, Staab, and Studer}]{sure2004knowledge}
\bibinfo{author}{Y.~Sure}, \bibinfo{author}{S.~Staab},
  \bibinfo{author}{R.~Studer}, \bibinfo{title}{On-To-Knowledge Methodology
  (OTKM)}, \bibinfo{publisher}{Springer Berlin Heidelberg},
  \bibinfo{address}{Berlin, Heidelberg}, \bibinfo{year}{2004}, pp.
  \bibinfo{pages}{117--132}.
%Type = Inproceedings
\bibitem[{De~Nicola et~al.(2005)De~Nicola, Missikoff, and
  Navigli}]{de2005proposal}
\bibinfo{author}{A.~De~Nicola}, \bibinfo{author}{M.~Missikoff},
  \bibinfo{author}{R.~Navigli},
\newblock \bibinfo{title}{{A proposal for a unified process for ontology
  building: UPON}},
\newblock in: \bibinfo{booktitle}{DEXA Proceedings}, \bibinfo{year}{2005}, pp.
  \bibinfo{pages}{655--664}.
%Type = Inbook
\bibitem[{Pinto et~al.(2006)Pinto, Staab, Tempich, and Sure}]{Pinto2006}
\bibinfo{author}{H.~S. Pinto}, \bibinfo{author}{S.~Staab},
  \bibinfo{author}{C.~Tempich}, \bibinfo{author}{Y.~Sure},
  \bibinfo{title}{{{Distributed Engineering of Ontologies (DILIGENT)}}},
  \bibinfo{publisher}{Springer Berlin Heidelberg}, \bibinfo{year}{2006}, pp.
  \bibinfo{pages}{303--322}.
%Type = Inbook
\bibitem[{Jarrar and Meersman(2009)}]{jarrar2009ontology}
\bibinfo{author}{M.~Jarrar}, \bibinfo{author}{R.~Meersman},
  \bibinfo{title}{Ontology Engineering -- The DOGMA Approach},
  \bibinfo{publisher}{Springer Berlin Heidelberg}, \bibinfo{year}{2009}, pp.
  \bibinfo{pages}{7--34}.
%Type = Inbook
\bibitem[{Su{\'a}rez-Figueroa et~al.(2012)Su{\'a}rez-Figueroa,
  G{\'o}mez-P{\'e}rez, and Fern{\'a}ndez-L{\'o}pez}]{suarez2011neon}
\bibinfo{author}{M.~C. Su{\'a}rez-Figueroa},
  \bibinfo{author}{A.~G{\'o}mez-P{\'e}rez},
  \bibinfo{author}{M.~Fern{\'a}ndez-L{\'o}pez}, \bibinfo{title}{{{The NeOn
  Methodology for Ontology Engineering}}}, \bibinfo{publisher}{Springer Berlin
  Heidelberg}, \bibinfo{year}{2012}, pp. \bibinfo{pages}{9--34}.
%Type = Article
\bibitem[{Iqbal et~al.(2013)Iqbal, Murad, Mustapha, Sharef
  et~al.}]{iqbal2013analysis}
\bibinfo{author}{R.~Iqbal}, \bibinfo{author}{M.~A.~A. Murad},
  \bibinfo{author}{A.~Mustapha}, \bibinfo{author}{N.~M. Sharef}, et~al.,
\newblock \bibinfo{title}{{An analysis of ontology engineering methodologies: A
  literature review}},
\newblock \bibinfo{journal}{Research journal of applied sciences, engineering
  and technology} \bibinfo{volume}{6} (\bibinfo{year}{2013})
  \bibinfo{pages}{2993--3000}.
%Type = Book
\bibitem[{Keet(2018)}]{keet2018introduction}
\bibinfo{author}{C.~M. Keet}, \bibinfo{title}{{An Introduction to Ontology
  Engineering}}, \bibinfo{publisher}{University of Cape Town},
  \bibinfo{year}{2018}.
%Type = Article
\bibitem[{Simperl and Luczak-R{\"o}sch(2014)}]{simperl2014collaborative}
\bibinfo{author}{E.~Simperl}, \bibinfo{author}{M.~Luczak-R{\"o}sch},
\newblock \bibinfo{title}{{Collaborative ontology engineering: a survey}},
\newblock \bibinfo{journal}{The Knowledge Engineering Review}
  \bibinfo{volume}{29} (\bibinfo{year}{2014}) \bibinfo{pages}{101--131}.
%Type = Phdthesis
\bibitem[{Vrande{\v{c}}i{\'c}(2010)}]{vrandevcic2009ontology}
\bibinfo{author}{Z.~Vrande{\v{c}}i{\'c}}, \bibinfo{title}{{Ontology
  Evaluation}}, Ph.D. thesis, {Karlsruhe Institute of Technology},
  \bibinfo{year}{2010}.
%Type = Article
\bibitem[{Guarino and Welty(2002)}]{ontoclean}
\bibinfo{author}{N.~Guarino}, \bibinfo{author}{C.~Welty},
\newblock \bibinfo{title}{Evaluating ontological decisions with ontoclean},
\newblock \bibinfo{journal}{Commun. ACM} \bibinfo{volume}{45}
  (\bibinfo{year}{2002}) \bibinfo{pages}{61–65}.
%Type = Inproceedings
\bibitem[{Roussey et~al.(2009)Roussey, Corcho, and
  Vilches-Bl\'{a}zquez}]{roussey2009catalogue}
\bibinfo{author}{C.~Roussey}, \bibinfo{author}{O.~Corcho},
  \bibinfo{author}{L.~M. Vilches-Bl\'{a}zquez},
\newblock \bibinfo{title}{A catalogue of {OWL} ontology antipatterns},
\newblock in: \bibinfo{booktitle}{Proceedings of the Fifth International
  Conference on Knowledge Capture}, K-CAP '09, \bibinfo{publisher}{Association
  for Computing Machinery}, \bibinfo{year}{2009}, p.
  \bibinfo{pages}{205–206}.
%Type = Inproceedings
\bibitem[{Karsai et~al.(2009)Karsai, Krahn, Pinkernell, Rumpe, Schneider, and
  Völkel}]{karsai2014design}
\bibinfo{author}{G.~Karsai}, \bibinfo{author}{H.~Krahn},
  \bibinfo{author}{C.~Pinkernell}, \bibinfo{author}{B.~Rumpe},
  \bibinfo{author}{M.~Schneider}, \bibinfo{author}{S.~Völkel},
\newblock \bibinfo{title}{{D}esign {G}uidelines for {D}omain {S}pecific
  {L}anguages},
\newblock in: \bibinfo{booktitle}{Proceedings of the 9th OOPSLA Workshop on
  Domain-Specific Modeling (DSM '09)}, volume \bibinfo{volume}{B-108},
  \bibinfo{year}{2009}, pp. \bibinfo{pages}{7--13}.
%Type = Book
\bibitem[{Fowler(2010)}]{fowler2010domain}
\bibinfo{author}{M.~Fowler}, \bibinfo{title}{Domain-specific languages},
  \bibinfo{publisher}{Pearson Education}, \bibinfo{year}{2010}.
%Type = Article
\bibitem[{Mernik et~al.(2005)Mernik, Heering, and Sloane}]{mernik2005and}
\bibinfo{author}{M.~Mernik}, \bibinfo{author}{J.~Heering},
  \bibinfo{author}{A.~M. Sloane},
\newblock \bibinfo{title}{When and how to develop domain-specific languages},
\newblock \bibinfo{journal}{ACM Comput. Surv.} \bibinfo{volume}{37}
  (\bibinfo{year}{2005}) \bibinfo{pages}{316–344}.
%Type = Article
\bibitem[{Crapo and Moitra(2013)}]{crapo2013toward}
\bibinfo{author}{A.~Crapo}, \bibinfo{author}{A.~Moitra},
\newblock \bibinfo{title}{Toward a unified english-like representation of
  semantic models, data, and graph patterns for subject matter experts},
\newblock \bibinfo{journal}{International Journal of Semantic Computing}
  \bibinfo{volume}{7} (\bibinfo{year}{2013}) \bibinfo{pages}{215--236}.
%Type = Incollection
\bibitem[{Gangemi and Presutti(2009)}]{gangemi2009ontology}
\bibinfo{author}{A.~Gangemi}, \bibinfo{author}{V.~Presutti},
\newblock \bibinfo{title}{{Ontology Design Patterns}},
\newblock in: \bibinfo{booktitle}{Handbook on Ontologies},
  \bibinfo{publisher}{Springer}, \bibinfo{year}{2009}, pp.
  \bibinfo{pages}{221--243}.
%Type = Book
\bibitem[{Ga{\v{s}}evi{\'c} et~al.(2006)Ga{\v{s}}evi{\'c}, Djuri{\'c}, and
  Deved{\v{z}}i{\'c}}]{gavsevic2006model}
\bibinfo{author}{D.~Ga{\v{s}}evi{\'c}}, \bibinfo{author}{D.~Djuri{\'c}},
  \bibinfo{author}{V.~Deved{\v{z}}i{\'c}}, \bibinfo{title}{Model driven
  architecture and ontology development}, volume~\bibinfo{volume}{10},
  \bibinfo{publisher}{Springer}, \bibinfo{year}{2006}.
%Type = Inproceedings
\bibitem[{Pan et~al.(2006)Pan, Xie, Ma, Yang, Qiu, and Lee}]{pan2006model}
\bibinfo{author}{Y.~Pan}, \bibinfo{author}{G.~Xie}, \bibinfo{author}{L.~Ma},
  \bibinfo{author}{Y.~Yang}, \bibinfo{author}{Z.~Qiu},
  \bibinfo{author}{J.~Lee},
\newblock \bibinfo{title}{{Model-Driven Ontology Engineering}},
\newblock in: \bibinfo{booktitle}{Journal on Data Semantics VII},
  \bibinfo{organization}{Springer}, \bibinfo{year}{2006}, pp.
  \bibinfo{pages}{57--78}.
%Type = Article
\bibitem[{Cranefield and Pan(2007)}]{cranefield2007bridging}
\bibinfo{author}{S.~Cranefield}, \bibinfo{author}{J.~Pan},
\newblock \bibinfo{title}{Bridging the gap between the model-driven
  architecture and ontology engineering},
\newblock \bibinfo{journal}{International Journal of Human-Computer Studies}
  \bibinfo{volume}{65} (\bibinfo{year}{2007}) \bibinfo{pages}{595--609}.
%Type = Inproceedings
\bibitem[{Lupp et~al.(2020)Lupp, Hodkiewicz, and
  Skj{\ae}veland}]{lupp2020template}
\bibinfo{author}{D.~P. Lupp}, \bibinfo{author}{M.~Hodkiewicz},
  \bibinfo{author}{M.~G. Skj{\ae}veland},
\newblock \bibinfo{title}{{Template libraries for industrial asset maintenance:
  A methodology for scalable and maintainable ontologies}},
\newblock in: \bibinfo{booktitle}{CEUR Workshop Proceedings}, volume
  \bibinfo{volume}{2757}, \bibinfo{organization}{Technical University of
  Aachen}, \bibinfo{year}{2020}, pp. \bibinfo{pages}{49--64}.
%Type = Inproceedings
\bibitem[{Gruninger(1995)}]{gruninger1995methodology}
\bibinfo{author}{M.~Gruninger},
\newblock \bibinfo{title}{{Methodology for the design and evaluation of
  ontologies}},
\newblock in: \bibinfo{booktitle}{Proceedings IJCAI'95, Workshop on Basic
  Ontological Issues in Knowledge Sharing}, \bibinfo{year}{1995}, pp.
  \bibinfo{pages}{1--10}.
%Type = Inproceedings
\bibitem[{Presutti et~al.(2009)Presutti, Daga, Gangemi, and
  Blomqvist}]{presutti2009extreme}
\bibinfo{author}{V.~Presutti}, \bibinfo{author}{E.~Daga},
  \bibinfo{author}{A.~Gangemi}, \bibinfo{author}{E.~Blomqvist},
\newblock \bibinfo{title}{{eXtreme design with content ontology design
  patterns}},
\newblock in: \bibinfo{booktitle}{Proceedings Workshop on Ontology Patterns},
  \bibinfo{year}{2009}, pp. \bibinfo{pages}{83--97}.
%Type = Article
\bibitem[{Chen(1976)}]{10.1145/320434.320440}
\bibinfo{author}{P.~P.-S. Chen},
\newblock \bibinfo{title}{The entity-relationship model—toward a unified view
  of data},
\newblock \bibinfo{journal}{ACM Trans. Database Syst.} \bibinfo{volume}{1}
  (\bibinfo{year}{1976}) \bibinfo{pages}{9–36}.
%Type = Misc
\bibitem[{Noy et~al.(2001)Noy, McGuinness et~al.}]{noy2001ontology}
\bibinfo{author}{N.~F. Noy}, \bibinfo{author}{D.~L. McGuinness}, et~al.,
  \bibinfo{title}{{Ontology development 101: A guide to creating your first
  ontology}}, \bibinfo{year}{2001}.
%Type = Article
\bibitem[{Booshehri et~al.(2021)Booshehri, Emele, Flügel, and et.
  al}]{BOOSHEHRI2021100074}
\bibinfo{author}{M.~Booshehri}, \bibinfo{author}{L.~Emele},
  \bibinfo{author}{S.~Flügel}, \bibinfo{author}{et. al},
\newblock \bibinfo{title}{{Introducing the Open Energy Ontology: Enhancing data
  interpretation and interfacing in energy systems analysis}},
\newblock \bibinfo{journal}{Energy and AI} \bibinfo{volume}{5}
  (\bibinfo{year}{2021}).
%Type = Inproceedings
\bibitem[{Skj{\ae}veland et~al.(2018)Skj{\ae}veland, Lupp, Karlsen, and
  Forssell}]{skjaeveland2018practical}
\bibinfo{author}{M.~G. Skj{\ae}veland}, \bibinfo{author}{D.~P. Lupp},
  \bibinfo{author}{L.~H. Karlsen}, \bibinfo{author}{H.~Forssell},
\newblock \bibinfo{title}{{{Practical Ontology Pattern Instantiation,
  Discovery, and Maintenance with Reasonable Ontology Templates}}},
\newblock in: \bibinfo{booktitle}{The Semantic Web -- ISWC 2018},
  \bibinfo{year}{2018}, pp. \bibinfo{pages}{477--494}.

\end{thebibliography}

\end{document}